\documentclass[prl,aps,twocolumn]{revtex4}
\usepackage{graphicx}
\usepackage{amsmath}
\usepackage{bm}
\usepackage{feynmp}

\newcommand{\beq}{\begin{equation}}
\newcommand{\eeq}{\end{equation}}
\newcommand{\bea}{\begin{eqnarray}}
\newcommand{\eea}{\end{eqnarray}}
\newcommand{\bei}{\begin{itemize}}
\newcommand{\eei}{\end{itemize}}
\newcommand{\etal}{{\em et al.}}

\newcommand{\br}{{\bf{r}}}

\DeclareMathOperator{\sgn}{sgn}
\DeclareMathOperator{\tanhShort}{th}

\def\x{x}
\def\y{y}
\def\X{X}
\def\Y{Y}
\def\p{k}
\def\Z{Z}

\begin{document}

\title{Density waves and supersolidity in rapidly rotating atomic Fermi gases}

\author{G. M\"{o}ller and N. R. Cooper}
\affiliation{Theory of Condensed Matter Group, Cavendish Laboratory, J.~J.~Thomson Ave., Cambridge CB3~0HE, UK}

\begin{abstract}

  We study theoretically the low-temperature phases of a two-component atomic Fermi gas
  with attractive $s$-wave interactions under conditions of rapid rotation.
  We find that, in the extreme quantum limit, when all particles occupy the
  lowest Landau level, the normal state is unstable to the formation of
  ``charge'' density wave (CDW) order.  At lower rotation rates, when many
  Landau levels are occupied, we show that the low-temperature phases can
  be supersolids, involving both CDW and superconducting order.

\end{abstract}
\date{April 29, 2007}
\pacs{
}
\maketitle

The experimental achievement of condensation of pairs of atoms in
two-component Fermi gases with resonant $s$-wave interactions
\cite{Jin04,Zwierlein04,Grimm04,Hulet04} has allowed studies of interacting
Fermi systems in regimes not accessible in solid-state systems: notably the
transition region between weak and strong interactions (BCS to BEC
crossover), and regimes of large density imbalance between the two species.
The ability to rotate the gases, revealing a lattice of quantized vortices
\cite{Zwierlein}, has provided an important diagnostic of superfluidity in these phase-coherent
condensates.

A very interesting regime arises in atomic Fermi gases under conditions of
rapid rotation (high vortex density). Noting the analogy between rotation
and magnetic field in a superconductor, one might anticipate the BCS phase to
revert to a normal state above a critical rotation frequency (analogous to
$H_{c2}$ in superconductivity), as predicted by BCS theory within a
semiclassical approximation \cite{Gorkov5960,Helfand66-II,Gurarie}.  Yet,
going beyond this approximation to include Landau level (LL) structure, one
finds that the normal phase can be unstable to ordered phases involving
high-field superconductivity (SC) \cite{Gruenberg}, ``charge'' density wave
(CDW), or spin-density wave (SDW) order \cite{Celli}. 

In this paper, we investigate the low-temperature phases of a two-component
atomic Fermi gas with attractive $s$-wave interactions under conditions
of rapid rotation.  The regime of interest for atomic gases differs
substantially from regimes studied in solid state systems: the rotation does
not lead to any ``Zeeman'' splitting which might suppress high-field SC order;
the short-range interactions allow density wave order to develop (this is
suppressed in solid state systems by Coulomb interactions).  We show that the
low-temperature phases of an atomic Fermi gas with attractive interactions
involve an interesting interplay between CDW and superconducting phases.  In
the extreme quantum limit, when only the lowest Landau level (LLL) is
occupied, we show that the system is unstable to CDW order along the rotation
axis. At lower rotation rates, we show that CDW and SC can coexist, leading to
``supersolid'' behaviour.

We study a rapidly rotating gas of two-species fermions, of equal densities,
in the uniform limit: the number of vortices is assumed large, so the rotation
frequency, $\Omega$, is close to the trap frequency, and the confinement
along the rotation axis is assumed weak.
In the rotating frame, the Coriolis force mimics a magnetic
field and leads to a Landau level
structure with cyclotron frequency $\omega_c=2\Omega$. The single
particle states then have energies
$\epsilon_\nu=(2n+1)\hbar\Omega + \frac{\hbar^2k^2}{2m}$, where
$\nu=(n,\x,\p)$ stands for the LL-index $n$, the momentum in the Landau gauge
$x$ 
\footnote{
For later convenience, the internal LL quantum number is denoted
  $\x$ (or $\y$) and
coordinate positions as $(\X,\Y,\Z)$,
adopting the notations of Ref.~\cite{Yakovenko93}.
}, and the wavevector along the rotation axis $k$.
For a non-interacting gas with Fermi energy $\epsilon_F$, the $n^{\rm th}$
Landau level has a 1D Fermi surface with Fermi momentum
$\hbar\p_{Fn}=[2m(\epsilon_F-(2n+1)\hbar\Omega)]^{1/2}$ and kinetic
energy relative to the bottom of the band $\epsilon_{Fn}=\hbar^2k_{Fn}^2/2m$.
We describe the instabilities of these Fermi surfaces arising from weak
interactions.  (We focus on results for attractive interactions, 
but also report on the repulsive case.)

First, we analyze the effect of rotation on the SC phase, applying BCS theory
in the presence of Landau level structure \cite{Gorkov5960,Helfand66-II,
  Helfand66-III,Gruenberg,Rieck,Rasolt}.  For contact interactions, the gap
equation requires regularisation at high energies.  For solid state systems,
the Debye frequency provides a natural cut-off for phonon-mediated
attractive interactions. In a cold atomic gas, a natural regularization arises
from the (small) lengthscale of the interparticle forces.  Using a two-channel
model for the Feshbach interaction, this lengthscale enters as the size of the
``closed channel'' boson (see e.g.
\cite{Ho,Gurarie,GurarieReview}) and can be taken to zero with the
introduction of appropriate counterterms. Following Ref.~\cite{Ho}, the
parameters of the model are the boson energy
$\epsilon_{\rm B}=2\hbar\Omega+\delta+C$ and the coupling $\alpha S_{\nu\nu'}$
between a closed channel boson and fermions with quantum numbers $\nu$ and
$\nu'$.  Here, $\delta$ is the physical detuning of the bosons and $C$ a
counterterm which is set to cancel the boson self energy $\Sigma(\omega\to
0)=\alpha^2\sum_{\nu\nu'}|S_{\nu\nu'}|^2/(\epsilon_\nu+\epsilon_{\nu'})$, such
that the model reproduces the scattering properties at low energy
and $\Omega\to 0$ \cite{Ho}\footnote{Eq.~(4) of Ref.~\cite{Ho} for $Q_{\nu\nu'}$
  is incorrect. 
  In our notation, the correct result is ($L_y=L_z=1$)
$$
S_{\nu\nu'} = c_X \delta(\p+\p') \frac{(-1)^n 2^{-N}}{\sqrt{\sqrt{2\pi}\,n!n'! \ell_0}}
H_N(\ell_0\sqrt{2}\bar\x) e^{-(\ell_0\bar \x)^2},
$$
with $\bar\x=(\x-\x')/2$, $N=n+n'$, and $\int dX/(2\pi) |c_X|^2=1$. 
}.  The physical scattering parameters
are related via $-\alpha^2/\delta = 4\pi\hbar^2 a_s/m
\equiv g$.  Treating the ensuing two-channel Hamiltonian within mean field,
and assuming a wide Feshbach resonance, yields the linearized gap equation \cite{Ho}
\begin{align}
\label{eq:GapEquation}
\frac{1}{-a_s} = {\hbar\Omega}
\!\! \sum_{n,n'=0}^{\infty} \!\! B^{n'}_{n} \!\!
\int \!\frac{dk}{2\pi}\!\left[
\frac{\tanhShort \frac{\xi_\nu}{2k_BT} + \tanhShort \frac{\xi_{\nu'}}{2k_BT}}{\xi_\nu+\xi_{\nu'}}
-\frac{2}{\epsilon_\nu\!+\!\epsilon_{\nu'}\!}\right]
\end{align}
with $B^{n'}_{n} =\binom{n+n'}{n} 2^{-n-n'}$, $\xi_\nu=\epsilon_\nu-\mu$, and the magnetic
length $\ell_0\equiv(\hbar/2m\Omega)^{1/2}$. The solutions to (\ref{eq:GapEquation}) determine
the critical temperature $T_c$ for  superconductivity.

\begin{figure}[ttp]
\includegraphics[width=0.99\columnwidth]{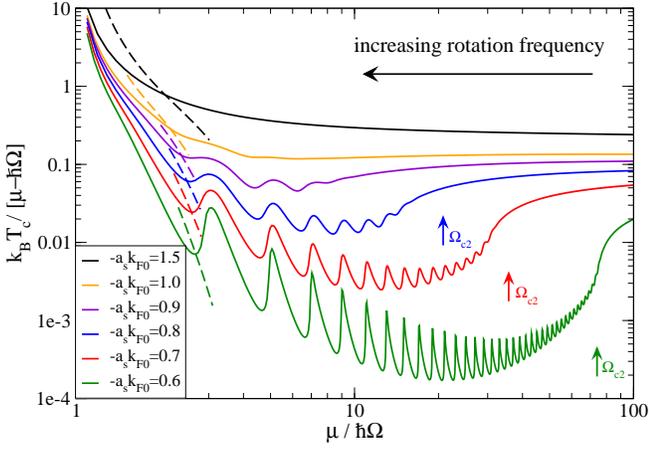}
\caption{ \label{fig:gaps} (color online) The critical temperature calculated
  within BCS mean-field, for rotation frequency $\Omega$ and chemical
  potential $\mu$. Arrows indicate critical frequencies $\Omega_{c2}$ to the left of which 
  $T_c$ vanishes in the semiclassical approximation \cite{Helfand66-II}. 
  Dashed lines show $T_c$ of the CDW state in the LLL as obtained in the 
  parquet approximation.}
\end{figure}

Within a semiclassical approximation to (\ref{eq:GapEquation}), $T_c$ vanishes
for $\hbar\Omega \gtrsim \Delta^2/\mu$ ($\Delta$ is the zero field
gap)\cite{Gorkov5960,Helfand66-II,Gurarie}\footnote{At $T=0$, the kinetic energy
of the superfluid flow in the vortex lattice (in the rotating frame) is
$\sim \hbar \Omega$ per particle, while the condensation energy is
$\sim\Delta^2/\mu$ per particle.}.  The full gap equation (1) admits solutions
even in this regime.
Then, when $T_c$ is small, the dominant contributions arise from
integrating the `diagonal' terms ($n=n'$)
\cite{Rieck}, which diverge logarithmically at low $T$ for occupied LL's.
Provided $k_BT_c \ll [\mu-\hbar\Omega(2n_\text{max}+1)]$, the off-diagonal
terms ($n\neq n'$) can be neglected, and one finds \beq
\label{eq:transitionT}
T_c \sim \eta \frac{\hbar\Omega}{k_B}  \exp\left\{- 
\frac{2\pi}{-a_s k_{F0}} G(\eta)^{-1} \right\}
\eeq
where $\eta\equiv(\mu-\hbar\Omega)/(2\hbar\Omega)$, 
$n_\text{max}=\lfloor \eta \rfloor$, and
\beq
\label{eq:GofEta}
G(\eta) \equiv \frac{1}{\eta}\sum_{n=0}^{n_\text{max}} \frac{(2n)!}{(2^n
  n!)^2} \left(1-\frac{n}{\eta}\right)^{-\frac{1}{2}}\,.  
\eeq 
The critical temperature (\ref{eq:transitionT}) is a strongly oscillating
function of $\mu/\hbar\Omega$, with a peak each time a LL depopulates
and $G(\eta)$ diverges.  The sharp peaks predicted by (\ref{eq:transitionT})
are rounded in a full of solution of (\ref{eq:GapEquation}) which is required
for strong-coupling. The evolution from weak to strong coupling is shown in
Fig.~\ref{fig:gaps}, which we have computed by solving (\ref{eq:GapEquation})
using a numerical root-finding routine. 

Consistent with previous studies of BCS theory in solid state systems
\cite{Gorkov5960,Helfand66-II, Helfand66-III,Gruenberg,Rieck,Rasolt} we find
that LL quantization leads to a stable SC state at \emph{any} value of the
field \cite{Gruenberg}.  
For $|a_s|k_F\lesssim 1$ the critical temperature has a minimum value 
$T_c^\text{min}(a_sk_F)$.
For temperatures $T\gtrsim T_c^\text{min}(a_sk_F)$, mean-field theory results 
predict a series of reentrant SC to normal transitions as the rotation rate increases.
Our results differ from those presented in Ref.~\cite{Ho}: the (reentrant)
superconductivity at rapid rotation was not found in that work; the critical
rotation frequency has an important temperature-dependence,
especially for strong coupling.

The superconducting phase 
competes with other ordered phases. To
determine the nature of the groundstate one must work {\it beyond} mean-field
theory.
We analyze the competition between SC and other ordered phases within a single
LL, for example when all particles occupy the LLL.
Owing to the quasi-1D dispersion within this LL,
any response function that connects opposite sides of the Fermi surface diverges 
at low temperatures as a power of
\beq
\label{eq:LogarithmicXi}
\xi = \frac{|g|}{(2\pi)^3 \hbar v_{Fn}
  \ell_0^2}\ln\left(\frac{\epsilon_{Fn}}{k_BT}\right) \, .  \eeq
Identical divergences occur in both particle-particle (p-p) and particle-hole
(p-h) diagrams \cite{Abrikosov}, and in diagrams of higher orders.  The
resulting ensemble of ``parquet'' diagrams, obtained by mutual insertion of
p-p and p-h-blocks into one another \cite{Brazowskii}, is most easily analyzed
in terms of a renormalization group (RG) approach \cite{Yakovenko93}. This
scheme has been applied to spinless electrons in a magnetic field
\cite{Yakovenko93}. We generalise this approach to a {\it two}-component
rotating atomic Fermi gas, and consider particles in the $n^{\rm th}$ LL.
Particles at the two Fermi points $\p=\pm\p_{Fn}$ are represented by separate
fermionic field operators $\hat a^{(\dagger)}$ and $\hat b^{(\dagger)}$.
States at the first Fermi point are expanded in terms of the LL
wavefunctions 
\cite{endnote24} 
\beq
\label{eq:LLWaveFunction}
\psi_{n\x\p}(\X,\Y,\Z)=\mathcal{N}_nH_n(\X-x)e^{i\x\Y+(\X-\x)^2/2+i\p\Z}, \eeq
with lengths measured in units of $\ell_0$, the
Hermite polynomials $H_n$ and normalization
$\mathcal{N}_n=(L_yL_z\pi^\frac{1}{2}2^n n!)^{-\frac{1}{2}}$.  
At the other Fermi point we use the transformed basis 
\beq
\tilde{\psi}_{n\y\p}(\X,\Y,\Z)=\frac{1}{N_\phi}\sum_x e^{-ixy} \psi_{n\x\p}(\X,\Y,\Z).
\eeq
For weak coupling, the kinetic energy can be linearized around the Fermi
points, $\pm \p_{Fn}$. The (logarithmically divergent) part of the contact interaction 
(amplitude $g$) describing scattering between opposite Fermi surfaces is
\begin{widetext}
\beq
\mathcal{H}_I=\frac{g}{L_xL_yL_z} \sum_{\mu\nu\sigma\rho} \sum_{\stackrel{\p_1,\p_2,\p_3}{x,x',y,y'}}
(\delta_{\mu\rho}\delta_{\nu\sigma}-\delta_{\mu\sigma}\delta_{\nu\rho})\, \gamma_0^{(n)}(x-x',y-y')\,
e^{i(xy'-x'y)} \hat a^\dagger_{n,x,\p_1,\mu} \hat b^\dagger_{n,y,\p_2,\nu} \hat b_{n,y',\p_3,\rho} 
\hat a_{n,x',\p_1+\p_2-\p_3,\sigma}
\label{eq:hi}
\eeq
\end{widetext}
The dependence on the LL index $n$ arises only in the form of the bare
interaction vertex, $ \gamma_0^{(n)}(\br) = e^{-\frac{1}{2}r^2} \left[L_n(
  r^2/2)\right]^2$, where we introduce $\br\equiv (x,y)$ and $L_n$
are the Laguerre polynomials.  The interaction $\mathcal{H}_I$ can be viewed
as two distinct vertices according to the way spin is conserved, and denoted
$\gamma_{1,2}$ in the usual notations for quasi-1D systems \cite{Solyom}. From
(\ref{eq:hi}), these vertices have the initial conditions \beq
\gamma_{1,2}^{(n)}(\br)|_{\xi=0}=\sgn(g) \gamma_0^{(n)}(\br).
\label{eq:initial}
\eeq
Renormalisation of the vertices $\gamma_{1,2}$ 
leads to corrections that can be expressed as
a power in $\xi$ (\ref{eq:LogarithmicXi}) \cite{Brazowskii}. 
The one-loop RG equations can be 
obtained by adapting the approach of Ref.~\cite{Yakovenko97} to include the LL
structure. We find
\begin{align}
\label{eq:RGequation1}
\frac{d\gamma_1}{d\xi} &= -2\, \gamma_1\ast\gamma_1 + 2\, \gamma_1\ast\gamma_2  
& - 2\, \gamma_1\otimes\gamma_2 \qquad\quad\;\;\;\\
\frac{d\gamma_2}{d\xi} &= \phantom{-}2\, \gamma_2\ast\gamma_2 & -  \gamma_1\otimes\gamma_1 - \gamma_2\otimes\gamma_2
\label{eq:RGequation2}
\end{align}
where the operations $\ast$ and $\otimes$ arise in p-h and p-p loops,
respectively, and are defined by 
\begin{fmffile}{fmfgraph2}
\beq
\parbox{9mm}
{\begin{fmfgraph}(20,20)
\fmftop{v1} \fmfbottom{v2} 
\fmf{fermion,right,tension=0.3}{v1,v2}
\fmf{scalar,right,tension=0.3}{v2,v1}
\fmfv{decor.shape=circle,decor.size=0.4w,decor.filled=empty}{v1}
\fmfv{decor.shape=circle,decor.size=0.4w,decor.filled=empty}{v2}
\end{fmfgraph}}
\equiv \gamma_i \ast \gamma_j \equiv \int d^2\br' \gamma_i(\br-\br') \gamma_j(\br').
\eeq
\end{fmffile}
\begin{fmffile}{fmfgraph3}
\beq
\parbox{9mm}
{\begin{fmfgraph}(20,20)
\fmfleft{v1} \fmfright{v2} 
\fmf{fermion,left,tension=0.3}{v1,v2}
\fmf{scalar,right,tension=0.3}{v1,v2}
\fmfv{decor.shape=circle,decor.size=0.4w,decor.filled=empty}{v1}
\fmfv{decor.shape=circle,decor.size=0.4w,decor.filled=empty}{v2}
\end{fmfgraph}}
\equiv \gamma_i \otimes \gamma_j \equiv \int d^2\br' 
\gamma_i(\br-\br') \gamma_j(\br') e^{-i\br\wedge\br'}.
\label{eq:conv2}
\eeq
\end{fmffile}
$\!\!\!$The  phase
factor in (\ref{eq:conv2}) is a consequence of the LL structure.

We have solved the RG equations (\ref{eq:RGequation1},\ref{eq:RGequation2})
with initial conditions (\ref{eq:initial}) for arbitrary Landau level index
$n$, using a standard numerical routine with $\gamma_{1,2}(|\br|)$ discretized
uniformly in $|\br|$. (The initial conditions for $\gamma_{1,2}$ are radially
symmetric, and this symmetry is preserved by the RG equations.)  To identify
instabilities, we calculate the renormalization of the response
functions \cite{Yakovenko97}.  The RG equations for the triangular vertices
$\mathcal{T}$ in the (singlet) SC, charge- and spin-density wave (SDW)
channels are given in our case by
\begin{align}
d_\xi \mathcal{T}_{\rm SSC} &= (-\gamma_1-\gamma_2)\otimes\mathcal{T}_{\rm SSC} \\
d_\xi \mathcal{T}_{\rm CDW} &= (-2\gamma_1+\gamma_2)\ast\mathcal{T}_{\rm CDW} \\
d_\xi \mathcal{T}_{\rm SDW} &= \gamma_2 \ast \mathcal{T}_{\rm SDW}.
\end{align}
Initial conditions for the triangular vertices can be chosen as
$\mathcal{T}_i|_{\xi=0}=\delta(\br)$, such that all Fourier components are
non-zero. We find the smallest value, $\xi_c$, at which a susceptibility
diverges: this indicates a transition into an ordered phase at a critical
temperature [see (\ref{eq:LogarithmicXi})] \beq
\label{eq:tc}
T_c \sim  \frac{\epsilon_{Fn}}{k_B} \exp{\left(-\frac{(2\pi)^3 \hbar v_{Fn} \ell_0^2}{|g|}\;\xi_c\right)}.
\eeq

In contrast to the full RG equations, the simplified equations describing only
p-h ladders can be solved analytically, and provide a useful reference point
for our numerical evaluation.  The solution for the p-h ladder discussed in
Ref.~\cite{Yakovenko93} can be generalized to arbitrary LL index $n$ and
yields a transition at a critical temperature which is independent of $n$ and
the sign of $g$, with $\xi_c=(2\pi)^{-1}$. [For $g<0$ ($g>0$) the transition
is to a CDW (SDW).] For p-p ladders, the problem can be solved analytically
for $n=0$, where the SC instability occurs for attractive interactions also at
$\xi_c=(2\pi)^{-1}$.  By restricting the SC gap equation in the presence of a
magnetic field to a single LL (see above and \cite{Rieck}), one can infer
$\xi_c(n)=(2^n n!)^2/[2\pi(2n)!]$, showing that SC order becomes weak as
$n\to\infty$.  These analytic results are reproduced by our numerical
approach, when restricted to include p-p or p-h diagrams only.
Note that for $n=0$ the CDW and SC instabilities have the same
critical temperature. Thus, mean-field theory cannot determine which
of these states will form the low-temperature phase. 

Our solution of the full RG equations
(\ref{eq:RGequation1},\ref{eq:RGequation2}) shows that, for attractive
interactions, CDW order is the dominant instability for all LLs. 
The critical temperature (\ref{eq:tc}) depends on the LL index, with exponents
summarized in Table~\ref{tab:Exponents}.  Thus, for the LLL 
$n=0$, the competition between the identical instabilities in p-p and p-h
channels [both at $\xi_c=(2\pi)^{-1}$] is decided to the advantage of CDW
order.  The order parameter diverges most strongly at zero in-plane momentum,
so the density-waves are aligned with the rotation axis.  Thus, the CDW phase
in the $n^{\rm th}$ LL 
involves a modulation of the particle density
along the rotation axis, with period $\lambda^{\rm CDW}_{n} = \pi/k_{Fn}$.
Within one period of the density wave the effective 2D particle density (in
that LL) is $n_{{\rm 2d},n} = 1/(\pi \ell_0^2)$, such that this LL
is fully occupied (its filling factor is $\nu_n \equiv n_{{\rm 2d},n}
2\pi\ell_0^2 = 2$).  Thus the CDW phase is fully gapped.  In the extreme
quantum limit, when $\hbar\Omega < \epsilon_F < 3\hbar\Omega$,
$k_{F0} = \pi^2 n \ell_0^2$, where $n$ is the 3D particle density, so the
period is $\lambda^{\rm CDW}_{0} = 1/(\pi n \ell_0^2)$.  
In Fig.~\ref{fig:gaps} we show the transition temperature into this CDW
in the LLL (dashed lines). (For repulsive
interactions, we find that SDW order is dominant
for all $n$. See Table~\ref{tab:Exponents}.)
\begin{table}
\begin{center}
\begin{tabular}{ccccccccccccc}
\hline
\hline
$n$ && 0 && 1 && 2 && 3 && 4  && $\infty$\\
\hline
$2\pi\xi_c|_{g<0}$ && 0.726(4) && 0.86(1) && 0.91(1) && 0.93(1) && 0.95(1) && 1 \\ %
$2\pi\xi_c|_{g>0}$ && 1.556(4) && 1.24(1) && 1.16(1) && 1.13(1) && 1.11(1) && 1 \\%
\hline\hline
\end{tabular}
\end{center} \caption{ \label{tab:Exponents} With dynamics restricted to a single Landau-level, 
the analysis of the parquet diagrams reveals a CDW instability for attractive interactions $g<0$, and 
a SDW instability for $g>0$.
While the CDW is enhanced by 
scattering in the p-p-channel, SDW order is weakened.
As $n\to\infty$ the critical values, $\xi_c$, converge to the result for p-h-ladders, $\xi_c = (2\pi)^{-1}$.}
\end{table}

Our results show that CDW order always prevails for
attractive contact interactions when dynamics
are restricted to a single LL.
However, at low rotation
rates, the groundstate 
is the BCS superconducting state
(with dilute vortices).
How does one reconcile these conclusions?
The answer lies in the coupling between LLs. 
Since the periods of the CDWs, $\lambda^{\rm CDW}_{n}$, differ between LLs, we
find that the CDW does not gain from inter-LL couplings: there are CDW
instabilities at the temperatures set by our calculations for individual LLs,
Table~\ref{tab:Exponents}.  On the other hand, a SC state {\it can} benefit
from coherence between LL's, as the Cooper pairs all have the same (zero)
momentum.  Thus, although SC within a single LL is less relevant than CDW, the
``Josephson'' coupling between LLs can stabilise a collective SC state.
That said, as the topmost LL, $n_{\rm max}$, depopulates our results show that
the CDW instability in this LL can occur at a higher temperature than the SC
state of the entire system.
In this case, the first instability (as $T$ is reduced) is to a CDW in the
Landau level $n_{\rm max}$, and one expects a second instability, at lower
$T$, to a SC state formed from the other Landau levels. (The loss of the
highest LL from the SC makes little difference to its condensation
energy.)  In this way, we predict a supersolid groundstate,
involving {\it both} CDW of the topmost LL and SC order in the lower LLs.
Ultimately, at sufficiently high rotation rate (or low particle density), when
all particles occupy the LLL, the groundstate 
is a CDW without superconducting order.

A striking consequence of our results is that for a rapidly rotating atomic
Fermi gas, there should appear spontaneous density wave order, with a period
$\lambda^{\rm CDW}_{n}$ that grows as the particle density in the topmost
Landau level decreases.  This can be a long lengthscale, so could be measured
in experiment directly by in situ absorption.  Clearly, the observation of the
density waves requires a trap with oscillator length $\ell_\parallel >
\lambda^{\rm CDW}_{n}$. For $\ell_\parallel < \lambda^{\rm CDW}_{n}$ there
will be a single period of the wave, leading to a quasi-2D regime with 2D
particle density in this LL equal to $n_{{\rm 2d},n}
=1/(\pi\ell_0^2)$. This (incompressible) filled LL will appear as a
step in the transverse density profile, as measured in-situ or 
in an expansion measurement \cite{readcooper}.

The results that we have presented are accurate far from the resonance on the
BCS side, where interactions are weak. We expect the qualitative behaviour to
survive as the resonance is approached. While the detailed energetics of both phases
cannot be relied upon for strong coupling, we find that SC is stabilized 
relative to CDW order for chemical potentials above the LLL as the coupling increases. 
Presumably, this leads to the suppression
of CDW states in any but the lowest LL as one approaches the resonance.
Furthermore, we note that the density-wave
state(s) we find on the BCS side of the resonance cannot evolve smoothly to the
BEC side. A CDW of atoms, with $\nu_{\rm atom}=2$ per period, could evolve, to
retain the same period, into a CDW of tightly bound molecules with $\nu_{\rm
  mol}=1/2$ per period \cite{cooper:feshbach}. However, there must be a phase
transition separating these two states, owing to the different edge structures
of the phases \cite{haldane}. Thus, in contrast to the SC phase at low
rotation rate, in the extreme quantum limit (at high rotation rate) 
tuning the interactions
across the Feshbach resonance must involve a phase transition.
\acknowledgments{We thank J. Chalker, M. Gunn, D. Haldane, W.
  Ketterle, and D. Khmelnitskii for helpful
  discussions. This work was supported by EPSRC (GR/S61263/01).}

\end{document}